\newcommand{\bD}{{\bf D}}
\newcommand{\psibar}{\overline{\psi}}

\def\GeV{{\rm GeV}}

\def\fm{{\rm fm}}

\def\vs{{\it vs.}}

\def\PS{\hbox{1P-1S}}
\def\SS{\hbox{2S-1S}}

\def\Bs{${\rm B_s}$}
\def\Ds{${\rm D_s}$}
\def\B{${\rm B}$}
\def\D{${\rm D}$}

\def\JournalRef#1#2#3#4{{\em #1}\/ {\bf #2} (#3) #4 }

\def\NuclPhys{Nucl.\ Phys.}

\def\LAT#1#2#3{\count255=#1\advance\count255 by 1
\JournalRef{\NuclPhys}{B (Proc. Suppl.) #2}{19{\the\count255}}{#3}}
 
\def\FigRef#1{Fig.~\ref{#1}}
\def\TabRef#1{Table~\ref{#1}}

\def\BorderBox#1#2{\vbox{\hrule height #1%
                         \hbox{\vrule width #1%
                               #2%
                               \vrule width #1%
                              }%
                         \hrule height #1%
                        }%
                  }
 
\def\Vev#1{\left<#1\right>}

\def\simle{\mathrel{\rlap{\raise 0.511ex\hbox{$<$}}%
                         {\lower 0.511ex\hbox{$\sim$}}%
                }}
 
\def\simge{\mathrel{\rlap{\raise 0.511ex\hbox{$>$}}%
                         {\lower 0.511ex\hbox{$\sim$}}%
                }}

\def\TextStyleOver#1#2{%
        {\kern.05em\raise.3ex\hbox{\the\scriptfont0 $#1$}%
        \kern-.1em{\the\scriptfont0 /}%
        \kern-.1em\lower.3ex\hbox{\the\scriptfont0 $#2$}}}

\def\Err(#1){\pm #1}
\def\aErr(+#1-#2){\left(\vphantom{0}^{+#1}_{-#2}\right)}

\def\FigHFS[#1]{%
\begin{figure}[#1]
\vspace*{0pt}
\BorderBox{0pt}{
\epsfxsize=3.0in \epsfbox{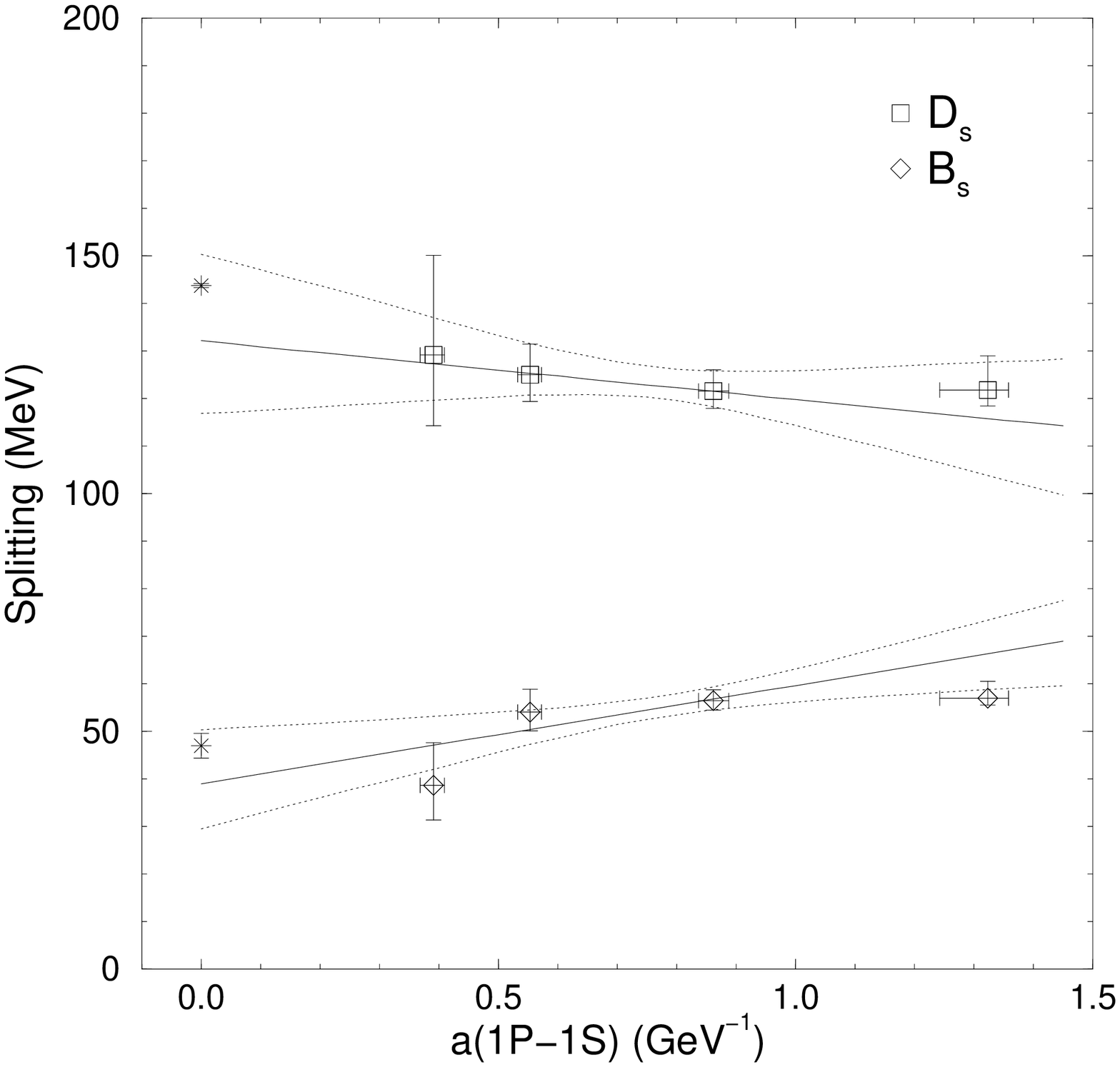}
}
\vspace*{0pt}
\caption{%
The  \Ds\ and  \Bs\ hyperfine splittings    \vs\ the lattice  spacing.
Linear extrapolations (solid)  with $68\%$ confidence  bounds (dashed)
are  shown.  Experimental splittings (bursts)  are indicated at $a=0$.}
\label{fig:HFS}
\end{figure}}

\def\FigRatios[#1]{%
\begin{figure}[#1]
\vspace*{0pt}
\BorderBox{0pt}{
\epsfxsize=3.0in \epsfbox{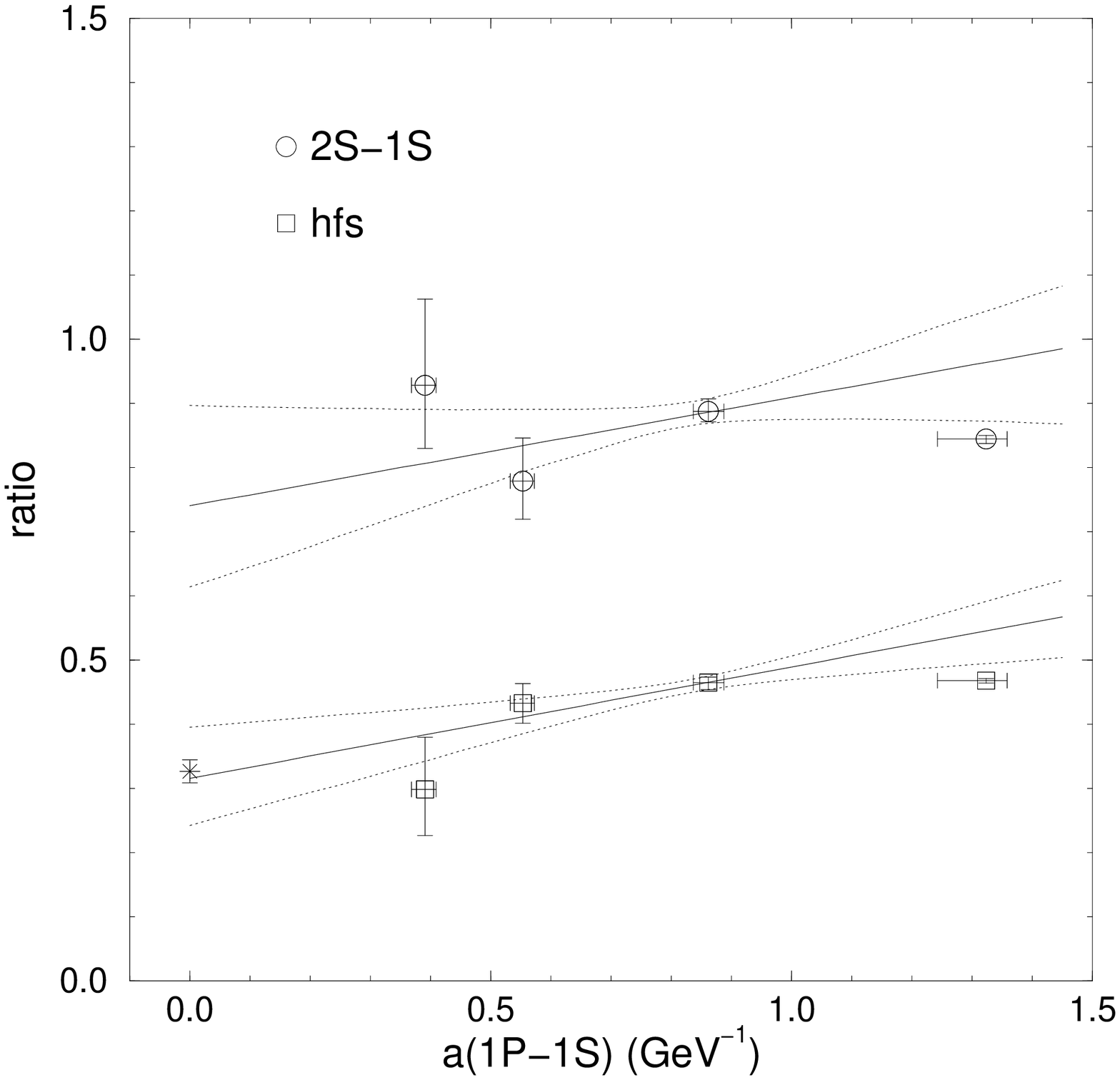}
}
\vspace*{0pt}
\caption{%
\Bs\  splittings divided   by  \Ds\ splittings.   Hyperfine  (hfs) and
spin-averaged 2S-1S splitting  ratios are shown.  The experimental hfs
ratio  (burst) is shown at $a=0$.    Linear extrapolations with errors
are shown.}
\label{fig:Ratios}
\end{figure}}

\def\FigSpect[#1]{%
\begin{figure}[#1]
\vspace*{0pt}
\BorderBox{0pt}{
\epsfxsize=3.0in \epsfbox{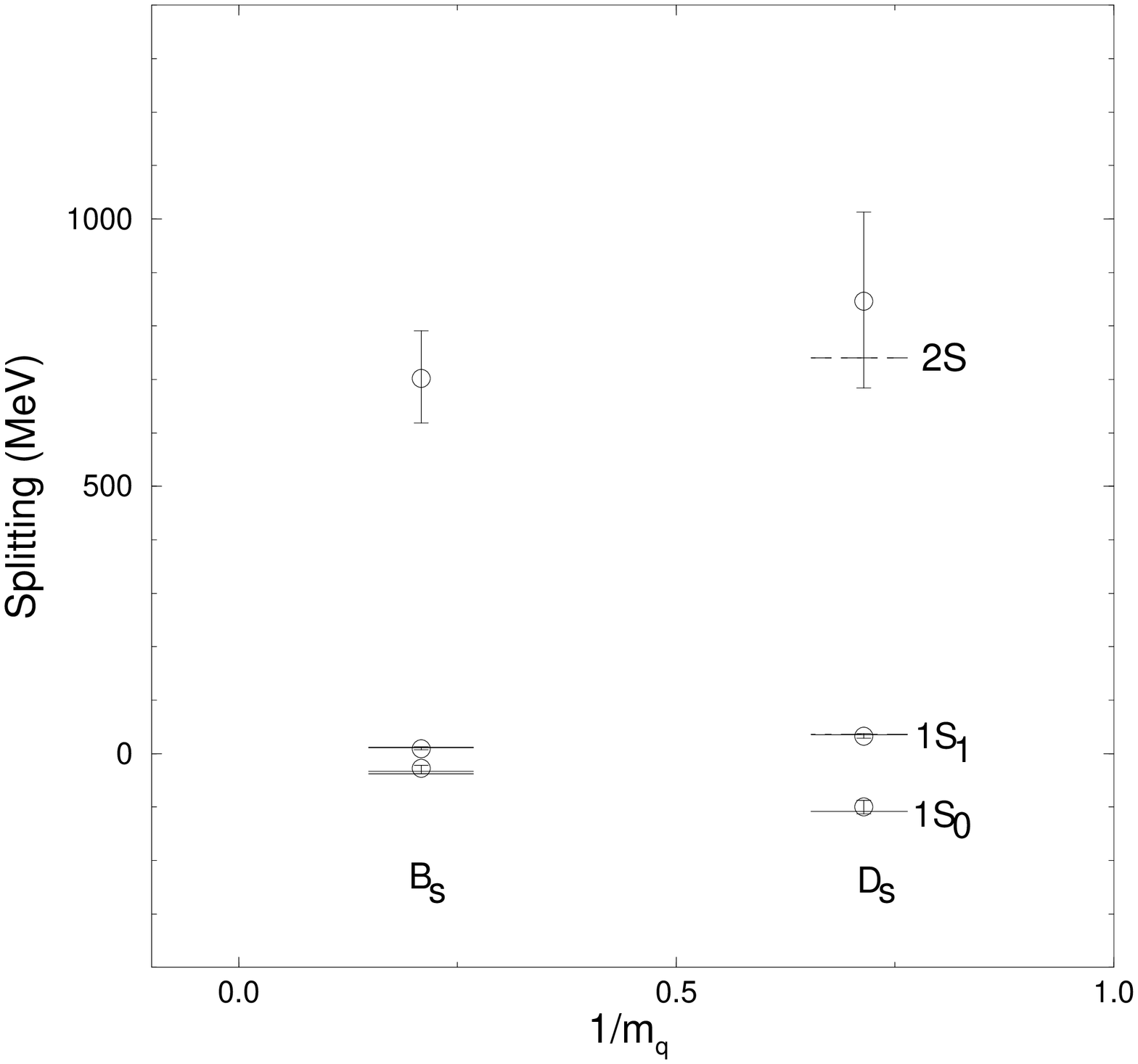}
}
\vspace*{0pt}
\caption{%
The \Ds\ and \Bs\ splittings  in  quenched QCD after extrapolation  to
zero  lattice spacing (octagons).   Experimental 1S levels (solid) are
indicated.  A potential model estimate of the \Ds\ spin-averaged \SS\
splitting (dashed) is also shown\protect\cite{Rosner}.  }
\label{fig:Spect}
\end{figure}}

\def\FigCharmoniumSpect[#1]{%
\begin{figure}[#1]
\vspace*{0pt}
\BorderBox{0pt}{
\epsfxsize=3.0in \epsfbox{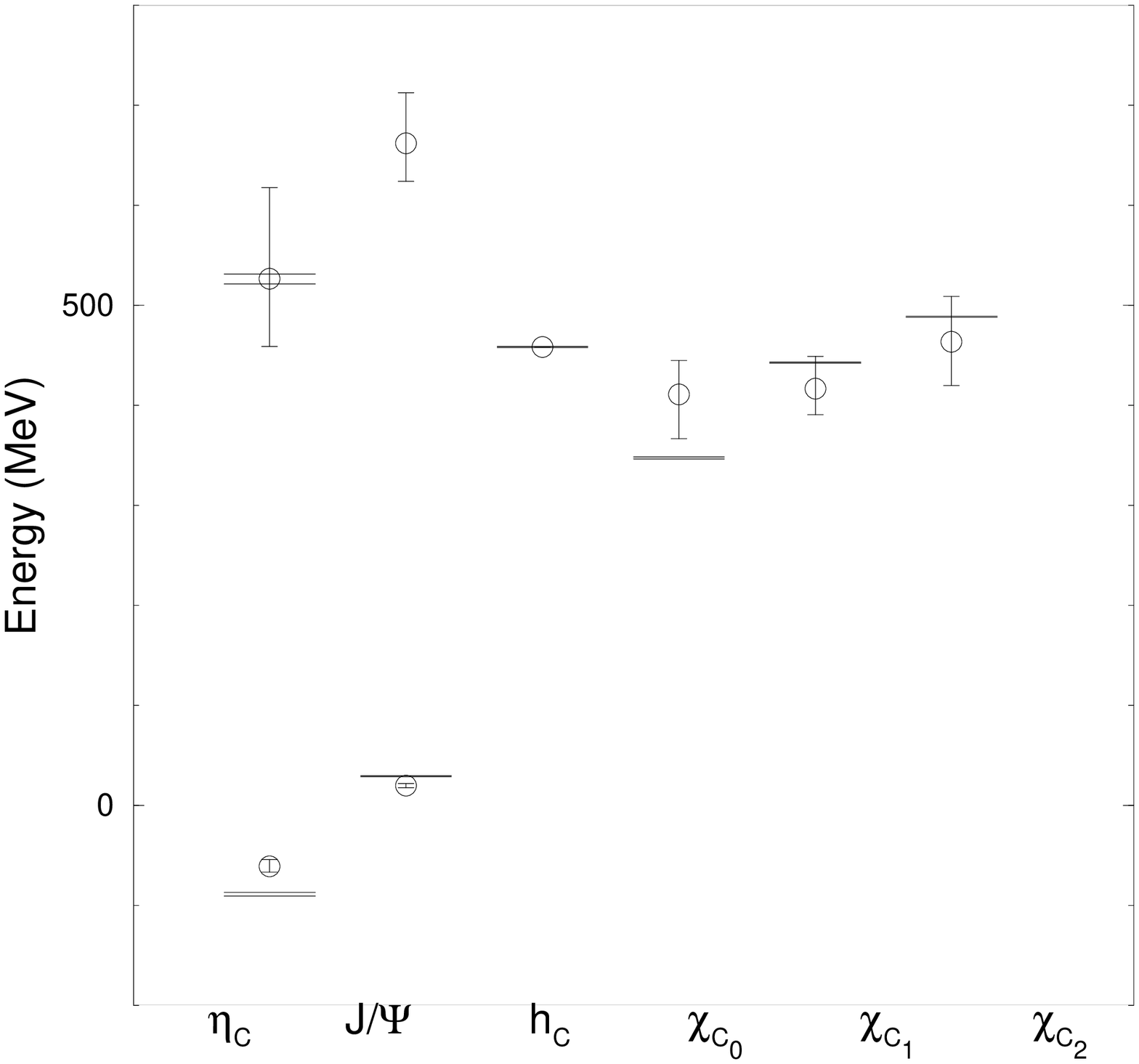}
}
\caption[charmonium]{%
Charmonium spectrum from the $\beta=6.1$ lattice.
}
\label{fig:charmonium}
\end{figure}}

\def\TabLAT[#1]{%
\begin{table*}[#1]
\setlength{\tabcolsep}{1.5pc}
\caption{Quenched gauge configuration ensembles.}
\label{tab:LAT}
\begin{tabular}{l|llll}
\hline
beta		&6.1	&5.9	&5.7	&5.5  \\
\# configurations&100	&350	&300	&500  \\
$c_{sw}=\Vev{{\rm plaq}}^{-3/4}$
		&1.46	&1.50	&1.57	&1.69 \\
$a^{-1}(\PS)\,\,(\GeV)$
		&$2.56\aErr(+16-12)$
		&$1.81(7)$
		&$1.16(3)$
		&$0.75\aErr(+8-2)$ \\
length $(\fm)$	&1.9	&1.7	&2.0	&2.1  \\
\hline
\end{tabular}
\end{table*}}

\documentstyle[twoside,fleqn,espcrc2,epsf]{article}

\title{Spectroscopy of \Bs\ and \Ds\ Mesons\thanks{
Talks presented by P.~Mackenzie and J.~Simone.
}}

\author{%
Paul B. Mackenzie$^a$,
Sin\'ead M.~Ryan
\address{Theoretical Physics Group,
Fermilab, P.O.\ Box 500, Batavia, Il 60510-0500, U.S.A.}
and 
James~N.~Simone%
\address{Loomis Laboratory of Physics, 1110 W.\ Green Street,
Urbana, Il 61801-3080, U.S.A.}%
}

\begin{document}

\begin{abstract}
We study \Bs\ and \Ds\ spectroscopy  in quenched lattice QCD using the
Fermilab approach to heavy quarks.   We obtain results at four lattice
spacings, $a$, using $O(a)$-improved  Wilson  quarks.
We compare and  contrast the various methods for  heavy quarks  on the
lattice, discussing discussing  which methods work best for  different
physical    systems and the   ease   with  which  calculations may  be
performed.

\end{abstract}

\maketitle

\section{ METHODS FOR HEAVY QUARKS ON THE LATTICE}
There are several ways of approximating heavy quarks in 
lattice QCD calculations, including Nonrelativistic 
QCD \cite{Cas86,Lep87,Lep92}  (NRQCD),
the static approximation \cite{Eic87,Hil90}, 
and the approach developed at Fermilab \cite{Kha97}
which takes the uncorrected Wilson action as its leading approximation
but adds correction operators which end up resembling those
of NRQCD rather than those of the standard Symanzik improvement
of the Wilson action \cite{Wil74}.

NRQCD is based on an expansion in nonrelativistic operators
(rotationally invariant but not  Lorentz invariant)
similar to that used in calculating relativistic corrections 
in the hydrogen atom.
It can be thought of as arising from a discretization of the 
action arising from a Foldy-Wouthuysen-Tani transformation
of the quark fields analogous to the one used in atomic physics.
\begin{equation}
\psi \rightarrow \exp(\theta D_i \gamma_i ) \psi
\end{equation}
leads to the action
\begin{eqnarray}\
D \! \! \! \! /+m &\rightarrow&  D_0	\gamma_0 +m 
	\nonumber \\ \label{NR}
  &+&  	 \frac{\bD^2}{2m} + \frac{(\bD^2)^2}{8m^3} + \ldots
\end{eqnarray}
The pole mass term does not affect the dynamics of nonrelativistic
systems and is conventionally dropped.

In $B$ physics, the simplest of all the methods can be used,
the static approximation.
This approximates the heavy quark propagator with a simple Wilson
line moving in the time direction,
giving an action corresponding to the first term in Eqn. \ref{NR}.
It is clearly most useful for the heaviest quarks.
It is not much used recently because it has a much worse 
signal to noise ratio than NRQCD, which is clear in retrospect 
but was not foreseen.

The third method can be thought of as arising from
 a partial FWT transformation:
\begin{equation}
\psi \rightarrow \exp(\theta' D_i \gamma_i ) \psi,
\end{equation}
and
\begin{eqnarray}
D \! \! \! \! /+m &\rightarrow&  D_0	\gamma_0 +m + a_1  D_i \gamma_i 
	\nonumber \\
  &+&  a_2 \left( \frac{\bD^2}{2m} + \frac{(\bD^2)^2}{8m^3} + \ldots \right),
\label{partialFTW}
\end{eqnarray}
where $\theta'<\theta$, and $a_1$ and $a_2$ are between 0 and 1.
This appears to be a crazy thing to do, producing an action which
combines the  defects of the transformed and the untransformed 
actions, which indeed it does.
On the other hand, it turns out that this is the action we 
have been stuck with for a long time.
The Wilson action has just this form at leading order.
The $\psibar D^2 \psi$ term added to cure the doubling problem 
also contributes to the kinetic energy, as in Eqn.~\ref{partialFTW}.
Furthermore, perhaps surprisingly, the $a_1$ and $a_2$ implied by the 
Wilson action have the desired property that the nonrelativistic
$\psibar D^2 \psi$ term takes over automatically from
the Dirac style kinetic energy term $\psibar D \! \! \! \! /  \psi$
 as $\kappa \rightarrow 0$
in the Wilson action.
The Wilson action automatically turns into a nonrelativistic action
in the large mass limit.

The Wilson action does have the unwanted property that for large masses 
 the pole mass does not equal the kinetic mass 
governing the energy momentum relation 
\begin{equation}
\frac{1}{2M_{kin}} = \frac{dE}{d p^2}.
\end{equation}
In nonrelativistic systems, the pole mass does not affect the dynamics
and the kinetic mass governs the leading important term.
Therefore, the pole mass term is normally simply omitted from the action 
in NRQCD, although there is no harm (and no benefit) in including it.
Likewise, the Wilson action can be used without problems 
for nonrelativistic systems as long as $M_{kin}$ and not the pole mass
 is used to set the quark mass.
It is easy to find a Wilson style action which does satisfy 
$M_{pole} = M_{kin}$
by letting the hopping parameter for the time direction, $\kappa_t$,
differ from the one in the spatial directions, $\kappa_s$.

The Wilson action corrected up to ${\cal O} (a)$ and the action of NRQCD
both use one-hop time derivatives.
When the quarks are heavy, this is a requirement, since two-hop
time derivatives introduce new ghost states and imaginary energies
when $ma>1$.
Therefore, the Wilson action applied to heavy paths cannot
follow the conventional Symanzik program of using two-hop
corrections at ${\cal O}(a^2)$, but must follow the NRQCD style
of corrections which correct only the spatial interactions.
The existence of the Hamiltonian and the transfer matrix ensures that
this is possible.

The parameters of the action in this approach must have nontrivial
mass dependence for large masses, just as do those of NRQCD.
For large masses, the wave function normalization, the relation between
the mass and the hopping parameter, etc., are completely different
from their $m=0$ values.
When $m<1$ (in lattice units), this mass dependence may be expanded
in power series.
For the Wilson and Sheikholeslami-Wohlert actions \cite{She85},
this yields just the usual series of operators, with the same
coefficients:
\begin{eqnarray} 	\nonumber
{\cal L}& = &  m \psibar\psi + a_1 m^2 \psibar\psi + \ldots	\\ \nonumber
&+&\psibar D \! \! \! \! /  \psi+ b_1 m \psibar  D \! \! \! \! / \psi 
   + b_2 m^2  \psibar D \! \! \! \! / \psi+ \ldots \\       
&+& c_{sw} \psibar \sigma_{\mu\nu} F_{\mu\nu} \psi + 
   c_{2} m \psibar \sigma_{\mu\nu} F_{\mu\nu} \psi + \ldots 
\end{eqnarray}
Thus, while for $m>1$ the action becomes very similar to NRQCD
in its behavior, for $m<1$ it may be regarded as an all orders in m
resummation of the usual operators of the Wilson and SW actions.
For hadrons containing charmed quarks, it is  possible to do
calculations with the Wilson and SW actions even with the old interpretation
of the coefficients.
However, since $am \approx 5 a \Lambda_{QCD}$, the ability to
sum up the required series in $m$ exactly is likely to produce a much
faster approach to the continuum limit.

\section{HEAVY--LIGHT MESONS}

\TabLAT[tbh]
Here we use this approach to calculate the spectrum of
$B_s$ annd $D_s$ mesons.
The strong  interactions of a  meson  containing a single heavy  
quark simplify  in the infinite mass limit.   The heavy-quark spin and
the light-quark total angular momentum, $j$, are separately conserved.
States with  total  spin  $J=j\pm1/2$  form  degenerate-mass doublets.
Hence,  the hyperfine splitting between  the   vector, $J=1$, and  the
pseudoscalar, $J=0$, is zero.  Heavy  quark symmetry is only approximate
at  finite quark  mass so  the hyperfine splittings   for \D\  and \B\
mesons are proportional to $1/m_Q$.

The sensitivity of  the  hyperfine splitting to  the  heavy-quark mass
makes spectroscopy of  the \B\ and  \D\ mesons an important check on
the procedure used to determine the bare charm and bottom masses.  The
bare  masses  are important  inputs to our   programs to determine the
renormalized quark masses and weak matrix elements  of the \D\ and \B\
mesons\cite{Andreas97,Sinead97}.

Heavy-quark masses in   this study are  determined in  quarkonia.  The
bare  mass is  found by demanding   the experimental spin-averaged
meson mass  match  the   kinetic  mass determined  from   the  lattice
energy-momentum  dispersion relation\cite{Simone95}.  Quark masses may
also be determined by studying  heavy-light mesons.  These masses will
differ from the masses  in -onia since  lattice spacing errors  differ
for the   two  methods\cite{Kro97}.  Mass  errors  with  either method
however  lead  to bounded uncertainties  in  quantities such as hadron
mass   splittings    and  matrix  elements.    Moreover,    quark-mass
uncertainties  in these quantities  are  overcome by extrapolation  to
zero lattice spacing.

\section{CALCULATION DETAILS}

We  study  pseudoscalar and vector   mesons on the   four ensembles of
quenched Wilson glue  described in \TabRef{tab:LAT}.  Lattice spacings
are determined  using the charmonium   \PS\ splitting.  Spacings range
from $a=0.26$  to $a=0.080\,\fm$. The coarsest  lattice only provide a
check  upon    cutoff effects; results are    not   used for continuum
extrapolations.

We use  Sheikholeslami  Wohlert (SW)  quarks with  tree-level  tadpole
improvement.  The clover  coefficient, $c_{sw}$, is determined from the
average  plaquette  (see \TabRef{tab:LAT}).    Calculations   are done
directly at  the charm and   bottom  quark masses using.
   Results  are  fully $O(a)$ improved.

Bare strange-quark masses are  determined from the  light pseudoscalar
spectrum  using leading-order   chiral  perturbation theory   and  the
experimental $\pi$ and $K$ masses.
Meson masses are extracted from minimum $\chi^2$  multistate fits to a
matrix  of meson correlators   having Coulomb-gauge 1S- and 2S-smeared
sources and sinks.  The full covariance matrix is retained in fits.

\section{SPECTROSCOPY}

\FigHFS[t]

Fig.~\ref{fig:HFS} shows the \Bs\  and \Ds\ hyperfine splittings for
the lattices  in  \TabRef{tab:LAT}.  Results   from the three   finest
lattices are  extrapolated linearly to  $a=0$.  The \Ds\ slope  has an
error  bound consistent with zero.   Hence  errors from the charm mass
determination  are barely   discernible with present  statistics.  The
\Bs\ slope is more significant indicating larger  errors in the bottom
quark mass determination. Reducing the  lattice spacing decreases  the
\Bs\  splitting. This is consistent   with  an underestimation of  the
bottom quark mass in -onia\cite{Kro97}.

The extrapolated \Bs\ and  \Ds\  hyperfine splittings are   consistent
with  experimental values.  The   central   values however lie   below
experiment.  This may be an indication that the quenched \Bs\ and \Ds\
hyperfine splittings are    smaller than experiment.  Note    that the
quenched  hyperfine  splittings  in quarkonia  lie significantly below
experiment.  Higher statistics are necessary to  determine if there is
a similar quenching effect for heavy-light mesons.

\FigRatios[tbh]

The    ratio of the \Bs\ hyperfine    splitting to the \Ds\ splitting,
$R_{hfs}$, is proportional to $m_{c}/m_{b}$  according to heavy  quark
symmetry.  The ratio   is expected to be  less  sensitive to quenching
errors  than the  individual splittings.  The  linear extrapolation in
\FigRef{fig:Ratios}  yields $R_{hfs}=0.32\pm0.08$.  This compares well
with the experimental value   $0.327\pm0.018$.  This is  evidence that
our  procedure  for determining  the  bare quark  masses  leads to the
correct continuum result.

\FigSpect[tbh]

We   also study 2S  states.  Excited  state  statistical errors are an
order of magnitude larger  than  ground state errors.   Excited states
are also subject  to  larger systematic errors.  Combined  errors  are
comparable   to the   spin splittings.   Hence   we   report only  the
spin-averaged 2S splitting, $m(\SS)$.

According to heavy quark symmetry splitting $m(\SS)$ is constant up to
$O(1/m_Q)$   corrections.   We check  this   in \FigRef{fig:Ratios} by
computing   the ratio  of  \SS\  splittings for \Bs\     and \Ds.  The
deviation from unity is consistent with $O(1/m_Q)$ effects.

We  summarize  our continuum  \Bs\   and \Ds\ spectroscopy results  in
\FigRef{fig:Spect}.  Results  are plotted  in  relation to  $1/m_Q$ to
emphasize the behavior of the splittings in the heavy quark limit.  We
compare  to experimental values  for the 1S  states. A potential model
inspired  estimate for  $m(\SS)$ is also  shown\cite{Rosner}.  
Our lattice results are consistent with experiment.

\FigCharmoniumSpect[tbh]

 Fig.~\ref{fig:charmonium}  shows our most recent charmonium
spectrum, which we will use in our discussion in the next section.
It is mainly a statistical updating of our previous results,
except for the $\chi_c$ states.
These have been done using nonrelativistic operators for the
first time rather than quark antiquark bilinears.
This allows us to obtain a result for the $\chi_{c_2}$ for the first time.

\section{WHICH METHOD IS BEST WHERE?}

The correction operators of NRQCD are simple and easy to organize.
NRQCD quark propagators can be calculated almost instantaneously.
Heavy quark actions which start from the Wilson action have the 
advantage that they continue to work as the $a\rightarrow 0$ limit
is taken.  However, correcting them to ${\cal O}(a^2)$ or to ${\cal O}(v^4)$ 
is messier than for NRQCD.

The action of NRQCD is an expansion around $m=\infty$ and is
better and better behaved, the larger the quark mass.
It breaks down in two distinct ways as $1/m $  becomes larger.
When $1/(ma)$ gets larger than 1, perturbative calculability of 
coefficient functions breaks down due to factors of $1/(ma)$
in the diagrams.
When $\Lambda_{QCD}/m$ gets larger than 1, 
the convergence of the series of nonrelativistic operators breaks down.
This series is very convergent for the $b$ quark,
and NRQCD works very well in $B$ and $\Upsilon$ physics.
The $\psi$ system is reasonably nonrelativistic, too,
as shown by the success of potential models.  
NRQCD works reasonably well for this system, too,
although in some cases not as well as originally hoped.

For example,  a recent investigation \cite{Tro97}
of the effects of $v^6$ corrections in NRQCD 
studied their effect on the hyperfine splitting of charmonium.
In results of a few years ago,
the  $v^4$, $a^2$ improved NRQCD results lay much closer to the 
experimental answer than the ${\cal O} (a)$ improved relativistic
Fermilab results.
The interpretation seemed to be that the higher order corrections included
in the NRQCD calculations made these results much more
accurate than the less improved relativistic results.
However, last year Trottier calculated some of the $v^6$ terms which
contribute to the hyperfine splitting and found results which
made a big jump back down toward zero.
The actual situation seems to be that the $v^2$ expansion is very slow to
converge for some quantities in the charmonium system.
 The ability of the harder to improve relativistic 
formalism to recover a Dirac-like form of the action in the
$a\rightarrow 0$ limit is an advantage here.

For $b$ quark physics, the situation is somewhat reversed.
Here, the nonrelativistic expansion is well-converged after a few orders.
(Potential models estimate that $v^2$ is about 1/4 in the $J/\psi$
and about 1/10 in the $\Upsilon$.)
On the other hand, these corrections have to be included to achieve
high accuracy. 
One does not have the option of replacing them and letting
the relativistic $\psibar D \! \! \! \! /  \psi$
terms in the action of Ref. \cite{Kha97} take over
when $ma<1$.
This would require  an absurdly high lattice spacing.

A case where this plays a more significant role than had been expected 
 is in the quantity $M_\Upsilon - 2 M_B$. \cite{Col96}
In the quarkonium systems, the quark momentum $p \rightarrow m_b \alpha_s$
as $m_b\rightarrow \infty$.
Therefore, $(p a)^2$ errors and $(p/m)^2$ errors also go to infinity
in absolute terms, but approach fixed  percentages of leading order
physical quantities.
In heavy-light systems on the other hand, the $b$ quark's momentum
$p \sim \Lambda_{QCD}$ as $m_b\rightarrow \infty$.
$(p a)^2$ errors and $(p/m)^2$ errors due to the $b$ quark 
approach constants in absolute terms.

Of the pole and kinetic masses in the equation
\begin{equation}
E = M_{\it pole} + \frac{1}{2M_{\it kin}} p^2 + \ldots,
\end{equation}
$M_{\it kin}$ is more affected by $v^2$ errors
than $M_{\it pole}$ since it is
higher order.
For example, we can obtain the spin--averaged $2M_{HL}-M_{HH}$
splittings in the $D_s-\Upsilon$ and $B_s - \psi$ systems
from the  data  from Figs. 1 -- 4.
We obtain 0.81 GeV in the charm system, compared with 1.08 GeV
experimentally.
We obtain -1.68 GeV in the bottom system
compared with 1.35 GeV experimentally.
Kronfeld used potential models to show that
this was roughly what was to be expected. \cite{Kro97}
The analogous effect in NRQCD improved only to ${\cal O}(a)$ is about
15\%. \cite{NRQpri}

It should be stressed that as a percentage error,
no error in $B$ or $\Upsilon$ physics is blowing up as $m\rightarrow \infty$.
For example, the uncertainty induced in determinations of $f_B$
from the 20\% ambiguity of using $M_B$ or $M_\Upsilon$ to determine the $b$
quark mass is only of order a few per cent, as can be seen by
examining a graph of $f_B\sqrt m$ vs. $1/m$.
It is simply one of the $p^2$ errors known to be present in the ${\cal O}(a)$
improved theory.
On the other hand, it is probably reduced from $p^2_\Upsilon$
to $p^2_B$ by obtaining $m_b$ from $M_B$ rather than $M_\Upsilon$
in this case.
It has been argued many times that quarkonia are nice systems
to use for error analysis and parameter setting because of the
tools that are available for understanding the errors.
When those tools tell us that some errors in quarkonia are rather large,
that strategy
may sometimes have to be altered.

For pure $B$ meson physics (not considering the $\Upsilon$ system), 
the situation is probably not as bad for
``relativistic'' heavy quark actions.  
(Relativistic is in quotes because for $b$ systems,
relativistic $\psibar D \! \! \! \! /  \psi$ terms in the action
have almost no effect compared with the nonrelativistic 
$\psibar \bD^2 \psi$ terms.)
In these systems, $v^2$ correction terms are probably negligible
since $v^2 \sim (\Lambda_{QCD}/m_b)^2 < 1\%$.
It is therefore possible that an ${\cal O}(a)$
improved heavy quark theory such as the heavy SW action
may do an adequate job here.
Perturbative corrections to decay constants, etc., are more
annoying to calculate than for NRQCD or the $m=0$ Wilson theory, 
but by no means intractable.

\section*{ACKNOWLEDGMENTS}

This work  is done in  collaboration with A.\  El-Khadra, and
A.\ Kronfeld.     Calculations were performed on    the  ACPMAPS
supercomputer built and operated  by the Fermilab computing  division.
Fermilab is operated by  University Research Association, Inc.   under
contract with the U.S. Department of Energy.


\begin{thebibliography}{9}

\bibitem{Cas86}
W. E. Caswell and G. P. Lepage, Phys. Lett. {\bf 167B} (1986) 437.
\bibitem{Lep87} 
G. P. Lepage and B. A. Thacker,
Nucl. Phys. B Proc. Suppl. {\bf 4} (1987) 199;\\
B. A. Thacker and G. P. Lepage, Phys. Rev. {\bf D43} (1991) 196.
\bibitem{Lep92} 
G. P. Lepage, L. Magnea, C. Nakhleh, U. Magnea, and K. Hornbostel,
Phys. Rev. {\bf D46} (1992) 4052.
%

\bibitem{Eic87} 
E. Eichten, Nucl. Phys. B Proc. Suppl. {\bf 4} (1987) 170.
\bibitem{Hil90} 
E. Eichten and B. Hill, Phys. Lett. {\bf B243} (1990) 427.
%
\bibitem{Kha97} A. X. El-Khadra, A. S. Kronfeld, P. B.
    Mackenzie,     Phys. Rev. {\bf D55} (1997) 3933.
\bibitem{Wil74} 
K.G. Wilson, Phys. Rev. {\bf D10} (1974) 2445.
%

\bibitem{She85}
  B. Sheikholeslami and R. Wohlert, Nucl. Phys. {\bf B259} (1985) 572.
%
\bibitem{Andreas97}
A.\ Kronfeld, these proceedings.
%
\bibitem{Sinead97}
S.\ Ryan and J.\ Simone, these proceedings.
%
\bibitem{Simone95}
J.\ Simone, 
Nucl. Phys. B (Proc. Suppl.)
	{\bf 47} (1996) 17.
%
\bibitem{Kro97}
A.\ Kronfeld, 
Nucl. Phys. B (Proc. Suppl.)
	{\bf 53} (1997) 401.%
\bibitem{Rosner}
J.\ Rosner, EFI-95-02 (hep-ph/9501291).
%


\bibitem{Tro97}
H. D. Trottier,    Phys. Rev. {\bf D} 55 (1997) 6844.

\bibitem{Col96} S. Collins et al., Nucl. Phys. B (Proc. Suppl.)
	{\bf 47} (1996) 455.

\bibitem{NRQpri} NRQCD collaboration, private communication.

\bibitem{Has97} S. Aoki et al. (JLQCD Collaboration),
 Nucl. Phys. Proc. Suppl. {\bf 53} (1997) 355.


\end{thebibliography}
\end{document}